\documentclass[prl,aastex,amsmath,amssymb,twocolumn,widetext,floatfix,aps,showpacs]{revtex4-1} 

\usepackage{graphicx}
\usepackage{bm}
\usepackage{xcolor}
\definecolor{myblue}{rgb}{0.0, 0.0, 0.6}
\usepackage{hyperref}
\hypersetup{
  colorlinks = true,
  citecolor  = myblue,
  linkcolor  = myblue,
  urlcolor   = myblue
}

\begin{document}

\title{
  Precision Small Scattering Angle Measurements of Elastic Proton-Proton Single and Double Spin Analyzing Powers at the RHIC Hydrogen Jet Polarimeter
}%

\author{A.~A.~Poblaguev}\email{poblaguev@bnl.gov}
\author{A.~Zelenski}
\author{E.~Aschenauer} 
\author{G.~Atoian}
\author{K.~O.~Eyser}
\author{H.~Huang}
\author{Y.~Makdisi}
\author{W.~B.~Schmidke}
\affiliation{%
 Brookhaven National Laboratory, Upton, New York 11973, USA
}%
\author{I.~Alekseev}
\author{D.~Svirida}
\affiliation{%
 Alikhanov Institute for Theoretical and Experimental Physics, 117218, Moscow, Russia
}%
\author{N.~H.~Buttimore}
\affiliation{%
  School of Mathematics, Trinity College, Dublin 2, Ireland
}%

\date{\today}

\begin{abstract}
  The Polarized Atomic Hydrogen Gas Jet Target polarimeter  is employed by the Relativistic Heavy Ion Collider (RHIC) to measure the absolute polarization of each colliding proton beam. Polarimeter detectors and data acquisition were upgraded in 2015 to increase solid angle, energy range and energy resolution. These upgrades and advanced systematic error analysis along with improved beam intensity and polarization in RHIC runs 2015 ($E_\textrm{beam}$\,=\,$100$\,GeV) and 2017 ($255$\,GeV) allowed us to greatly reduce the statistical and systematic uncertainties for elastic spin asymmetries, $A_\textrm{N}(t)$ and $A_\textrm{NN}(t)$, in the Coulomb-nuclear interference  momentum transfer range $0.0013$\,$<$\,$-t$\,$<$\,$0.018$\,GeV$^2$. For the first time hadronic single spin-flip $r_5$ and double spin-flip $r_2$ amplitude parameters were reliably isolated at these energies and momentum transfers.  Measurements at two beam energies enable a separation of Pomeron and Regge pole contributions to $r_5(s)$ and $r_2(s)$, indicating that the spin component may persist at high energies.
\end{abstract}

\pacs{ %
24.70.+s, 
25.40.Cm, 
29.25.Pj.
}
 
\maketitle

{\em Introduction}.---%
Study of the spin-averaged elastic $pp$ hadronic amplitude at high energies has a more than 50 year history\,\cite{bib:PDG} and is continuing at the Large Hadron Collider. An essential contribution to this study relates to forward scattering for which the optical theorem and Coulomb-nuclear interference (CNI) provide an opportunity to separate the real and imaginary parts of an amplitude. Regge theory, based on the analyticity of a scattering amplitude, is a recognized method of understanding the energy dependence of amplitudes\,\cite{bib:Kaidalov}. 

An explanation of the unexpected discovery in the seventies  of a growing $pp$ cross section at high energies\,\cite{bib:Serpukhov} was found\,\cite{bib:Pomeron} in the Pomeron concept, which is now associated  with the exchange of nonperturbative QCD gluons\,\cite{bib:BFKL}. Currently, the Pomeron and  Regge pole picture of unpolarized elastic $pp$ scattering is commonly considered as well established in the $\sqrt{s}$\,=\,$5$\,GeV$\textendash13$\,TeV c.m. energy range\,\cite{bib:PDG}, though some new results, e.g. from the TOTEM experiment\,\cite{bib:TOTEM}, call for a revision\,\cite{bib:Odderon}. However, the accuracy of existing polarized high energy experimental data\,\cite{bib:FNAL,bib:HJET06,bib:HJET09,bib:STAR} was insufficient to identify a Pomeron contribution, if any, to the $pp$ spin-dependent amplitudes.

In this Letter, we report new measurements of the single spin $A_\textrm{N}(t)$ and double spin $A_\textrm{NN}(t)$ analyzing powers in  the small angle elastic collision of RHIC{\rq}s polarized proton beams with 
Polarized Atomic Hydrogen Gas Jet Target\,\cite{bib:ABS}(HJET) 
at $\sqrt{s}$\,=\,$13.76$ and $21.92$\,GeV. The precision has been significantly improved compared to previous
HJET 
publications\,\cite{bib:HJET06,bib:HJET09} and this has allowed us to not only isolate hadronic spin-flip amplitudes but also to incorporate spin dependence in a Regge pole analysis. It appears that forward elastic $pp$ scattering has nonvanishing single and double spin-flip hadronic amplitudes at high energy where the Pomeron dominates. The results of the analysis facilitate extrapolation of the measured $A_\textrm{N}(t)$ to a wide range of energies, essential for CNI polarimetry. Additional measurements at the  RHIC injection energy ($E_\textrm{beam}$\,=\,$24$\,GeV) might yield an improved Reggeon fit and the possibility\,\cite{bib:SPIN18} of experimentally resolving the Odderon issue\,\cite{bib:Odderon}.

 The HJET provides an absolute proton beam polarization measurement averaged across a beam. Typically, $\langle P_\textrm{beam}\rangle$\,$\sim$\,$55\pm2.0_\textrm{stat}\pm0.3_\textrm{syst})\%$\;\cite{bib:PSTP2017} for an 8-h RHIC store. The achieved accuracy satisfies the requirements of hadron polarimetry for planned and future accelerators such as the Electron Ion Collider (EIC)\,\cite{bib:EIC}. This work is based on the technique of high energy beam polarization measurement developed at RHIC.
The methodology can be recommended for EIC including a possible extension of it using other polarized nuclei such as $^3$He.

{\em HJET Polarimeter at RHIC.---}%
The HJET\,\cite{bib:ABS} acts like a fixed target that  measures absolute polarization of $24\textendash255$\,GeV proton beams at RHIC.
It consists of three main components: an atomic beam source, a Breit-Rabi polarimeter to measure atomic hydrogen polarization, and a recoil spectrometer to determine the beam and vertically polarized atomic hydrogen target (the jet)  spin correlated asymmetries of the detected recoil protons. Polarizations of both RHIC beams (alternating spin up/down bunches), so-called {\em{}blue} and {\em{}yellow}, are measured concurrently and continuously.

The jet density profile in the horizontal direction is well approximated by a Gaussian distribution ($\sigma_\textrm{jet}$\,$\approx$\,$2.6$\,mm), with $1.2$\,$\times$\,$10^{12}$\,atoms/cm$^2$ in the center. Since the RF-transition efficiency exceeds $99.9\%$, the polarization, $P_\textrm{jet}$\,$\approx$\,$0.96$, is defined by the strength ($1.2$\,kG) of the holding field magnet\,\cite{bib:HJET09}. The atomic hydrogen spin direction  is reversed every $5\textendash10$\,min.

\begin{figure}[t]
  \begin{center}
    \includegraphics[width=\columnwidth]{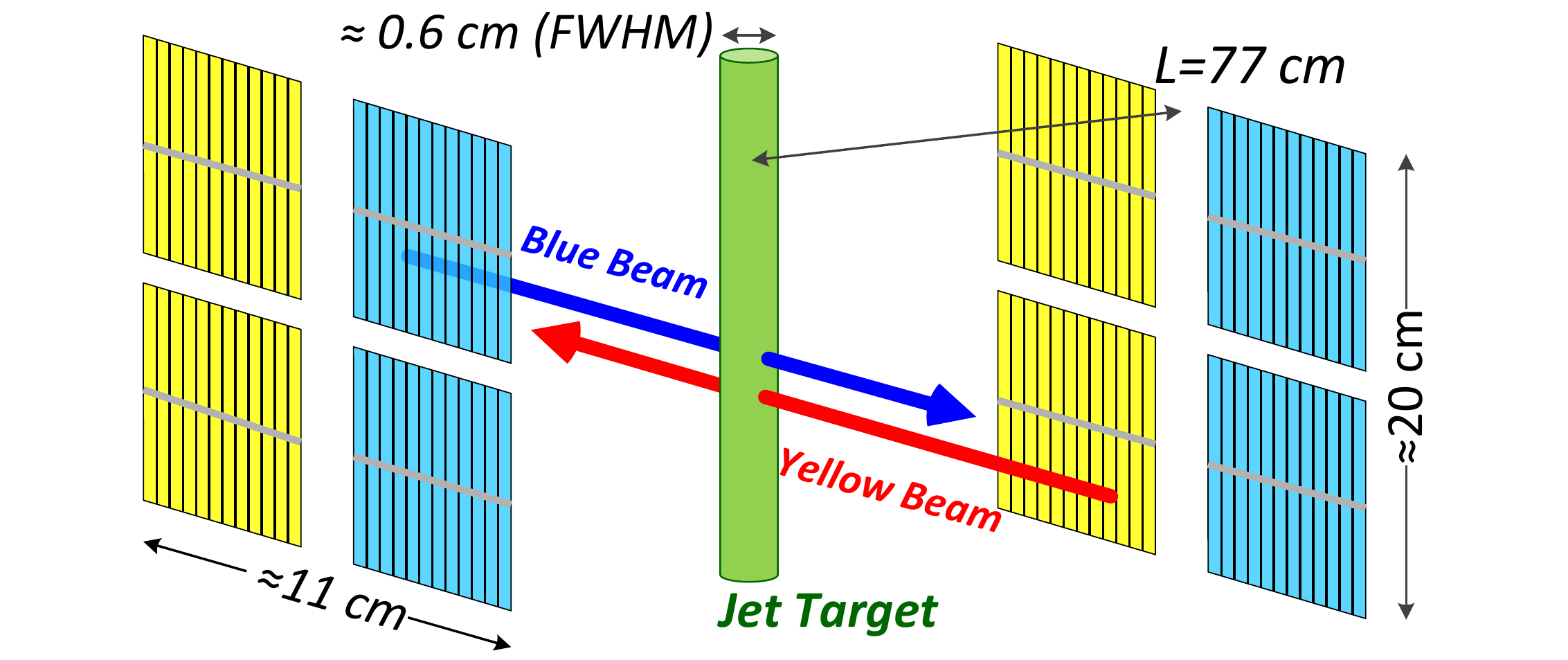}
    \caption{
      A schematic view of the HJET recoil spectrometer consisting of eight silicon detectors with 12 vertically oriented strips (readout channels) each. The distance between beams is $\sim$2\,mm.
    } \label{fig:HjetView}
  \end{center}
\end{figure}

The recoil spectrometer is sketched in Fig.\,\ref{fig:HjetView}. 
For elastic $pp$ scattering, the spectrometer geometry  allows us to detect recoil protons with kinetic energy up to $T_R$\,$\approx$\,$10\textendash11$\,MeV, i.e., to $-t$\,=\,$2m_pT_R$\,$\sim$\,$0.02$\,GeV$^2$. To reconstruct the kinetic energy of punch through protons ($T_R$\,$>$\,$7.8$\,MeV), signal waveform shape analysis was carried out. 

A detailed description of the HJET data analysis is given in Ref.\,\cite{bib:PSTP2017}. A crucial part of the analysis relates to an accurate determination of the background event rate in every Si detector as a function of  the measured energy and the spins of the jet and beam. Hence, to a subpercent level, spin effects were properly treated in the background subtraction.

{\em Spin correlated asymmetries.---}%
To measure the proton beam polarization, we studied the spin-correlated differential cross section\,\cite{bib:Convention,bib:Leader}
\begin{equation}
  \frac{d^2\sigma}{dtd\varphi}\!\propto\!
  \big[1\!+\!A_\textrm{N}(t)\sin\!{\varphi}\left(P_j\!+\!P_b\right)\text{+}A_\textrm{NN}(t)\sin^2\!{\varphi}P_jP_b\big]
\label{eq:phi}
\end{equation}
dependence on azimuthal angle $\varphi$. At HJET, $\sin{\varphi}$\,=\,$\pm1$ depending on right or left position of the Si detector relative to the beam. $P_{j,b}$ are the jet and beam polarizations, respectively. To determine analyzing powers $A_\textrm{N}(t)$ and $A_\textrm{NN}(t)$, the single spin (jet and beam) and double spin asymmetries
\begin{equation}
  a_\textrm{N}^\textrm{j}=A_\textrm{N}|P_{j}|,~~~ 
  a_\textrm{N}^\textrm{b}=A_\textrm{N}|P_{b}|,~~~ 
  a_\textrm{NN}=A_\textrm{NN}|P_jP_b|
\end{equation}
were derived\,\cite{bib:SPIN18} from the selected elastic event counts $N^{(\uparrow\downarrow)(+-)}_\textrm{RL}$ discriminated by the right/left ($\textrm{RL}$) detector location and the beam ($\uparrow\downarrow$) and jet ($+-$) spin directions.

For CNI elastic $pp$ scattering at high energies, the theoretical basis for an experimental parametrization of the analyzing powers was introduced in Refs.\,\cite{bib:KL,bib:BGL} and updated\,\cite{bib:BKLST} for the RHIC spin program. The analyzing powers can be written in terms of the anomalous magnetic moment of a proton  $\varkappa$\,=\,$1.793$, the unpolarized $pp$ scattering parameters $\rho(s)$ (forward real-to-imaginary amplitude ratio), $\sigma_\textrm{tot}(s)$ (total cross section), $B(s)$ (the nuclear slope)  and  hadronic single, $r_5$\,=\,$R_5+iI_5$, and double, $r_2$\,=\,$R_2+iI_2$, spin-flip amplitude parameters:
\begin{widetext}
\begin{eqnarray}
  \frac{m_p}{\sqrt{-t}}\,A_\textrm{N}(t) & = & \frac 
    {\left[\varkappa'(1-\rho'\delta_C)-2(I_5-\delta_C R_5)\right]\,t'_c/t -2\left(R_5 -\rho'I_5\right) }
    {\left(t_c/t\right)^2-2(\tilde{\rho}+\delta_C)\,t_c/t+1+\tilde{\rho}^2},
  \label{eq:AN}  \\ 
  A_\textrm{NN}(t) & = & \frac
    {-2(R_2 + \delta_CI_2)\,t'_c/t + 2(I_2 + \rho' R_2) - ( \rho'\varkappa'-4R_5)\,\varkappa' t_c/2m_p^2}
    {\left(t_c/t\right)^2-2(\tilde{\rho}+\delta_C)t_c/t+1+\tilde{\rho}^2}.
  \label{eq:ANN} 
\end{eqnarray}
\end{widetext}
In Ref.\,\cite{bib:BKLST}, terms $\varkappa'$, $\rho'$, $\tilde{\rho}$, and $t_c'$ in Eqs.\,(\ref{eq:AN})--(\ref{eq:ANN}) appeared as $\varkappa$, $\rho$, $\rho$, and  $t_c$, respectively. For the HJET measurements,  $-t_c$\,=\,$8\pi\alpha/\sigma_\textrm{tot}$\,$\approx$\,$0.0018$\,GeV$^2$  and the Coulomb phase is $\delta_C$\,=\,$-\alpha\ln\left|0.8905\left(B+8/\Lambda^2\right)t\right|$\,$\sim$\,$0.02$~\cite{bib:BKLST}.

Recently, it has been pointed out\,\cite{bib:Krelina} that Eqs.\,(\ref{eq:AN})--(\ref{eq:ANN}) were derived in Ref.\,\cite{bib:BKLST} with some simplifications. For the increased precision of the HJET measurements, corrections to $A_\textrm{N}(t)$ and $A_\textrm{NN}(t)$ should be applied. Some of them  have been outlined in Ref.\,\cite{bib:Corr}, in particular, {\em(i)} the difference between $pp$ electromagnetic and hadronic form factors and {\em(ii)} an additional term $\sim m_p^2/s$ in the single spin-flip electromagnetic amplitude. These corrections can be represented by the following substitutions:
\begin{eqnarray}
 t'_c      &=& t_c\times\left[1  + \left(r_p^2/3-B/2-\varkappa/2m_p^2\right)t\right],    \label{eq:tc'} \\
\rho'      &=& \rho + 
          \left(r_p^2/3-4/\Lambda^2-\varkappa/2m_p^2-\varkappa^2/4m_p^2\right)t_c\approx\rho,~~{} \label{eq:rho'} \\
\tilde{\rho}\phantom{'}     &=& \rho - \left(4/\Lambda^2-B/2\right)t_c,   \\
\varkappa' &=& \left(\varkappa-2m_p^2/s\right)/\left(1-\mu_pt/4m_p^2\right)  \label{eq:kappa'}
\end{eqnarray}
where $\Lambda^2$\,=\,$0.71$\,GeV$^2$, and $r_p$\,=\,$0.875$\,fm (CODATA\,\cite{bib:CODATA}) is a proton charge radius.

In most measurements of $\rho$, the $pp$  electromagnetic form factor $\mathcal{F}^\textrm{em}(t)$ was approximated in data analysis by  $\left(1-t/\Lambda^2\right)^{-4}$ derived from the electric form factor in dipole form\,\cite{bib:GD1966}. Therefore, the value of $\rho'-\rho$\:$\approx$\:$0.002$ should be interpreted as a systematic correction to be applied to the value of $\rho$ obtained from these experiments.  This correction might be essential for the Regge pole fit of the unpolarized data; however, it is completely negligible for this work.

The absorptive corrections to $\mathcal{F}^\textrm{em}(t)$, due to the initial and final state hadronic interactions between the colliding protons\,\cite{bib:Krelina}, are currently unavailable\,\cite{bib:Boris} and, consequently, are not included in the fits to the analyzing powers. However, if they effectively modify $\mathcal{F}^\textrm{em}$\:$\to$\:$\mathcal{F}^\textrm{em}$\,$\times$\,$\left[1+a(s)\,t/t_c\right]$ then the result of the fit using Eq.\,(\ref{eq:AN}) should be corrected\,\cite{bib:Corr}  by
\begin{equation}
\Delta_a R_5=a_\textrm{sf}\varkappa/2,~~~\Delta_a I_5=-a_\textrm{nf}\delta_C\varkappa/2\approx0
\end{equation}
where \lq\lq{sf}\rq\rq and \lq\lq{nf}\rq\rq denote the spin-flip and non-flip absorptive corrections, respectively.

{\em Analyzing power measurements at $\sqrt{s}$\,=\,$13.76$}\,GeV {\em and $\sqrt{s}$\,=\,$21.92$}\,GeV.---%
Here we analyze HJET data acquired in two RHIC proton-proton runs: Run\,15 ($100$\,GeV)\,\cite{bib:Run15} and Run\,17 ($255$\,GeV)\,\cite{bib:Run17}. About $2$\,$\times$\,$10^9$ elastic $pp$ events were selected at HJET in each run. 
In the data analysis, the values of $\sigma_\textrm{tot}(s)$ and $\rho(s)$ were taken from the $pp$ and $\bar{p}p$ data fit\,\cite{bib:Menon}. The slopes $B(s)$ were derived from Ref.\,\cite{bib:Bartenev}. 
The run specific conditions of the measurements can be briefly summarized as\\
\noindent{Run\,15:} 
$\sqrt{s}$\,=\,$13.76$\,GeV, $\rho$\,=\,$-0.079$,
$\sigma_\textrm{tot}$\,=\,$38.39$\,mb,
$B$\,=\,$11.2$\,GeV$^{-2}$, $P_\textrm{jet}^\textrm{eff}$\,=\,0.954;\\
\noindent{Run\,17:} 
$\sqrt{s}$\,=\,$21.92$\,GeV, $\rho$\,=\,$-0.009$,
$\sigma_\textrm{tot}$\,=\,$39.19$\,mb,
$B$\,=\,$11.6$\,GeV$^{-2}$,
$P_\textrm{jet}^\textrm{eff}$\,=\,$0.953$;\\
where $P_\textrm{jet}^\textrm{eff}$ is the effective jet polarization after systematic corrections.

\begin{figure}[t]
  \includegraphics[width=\columnwidth]{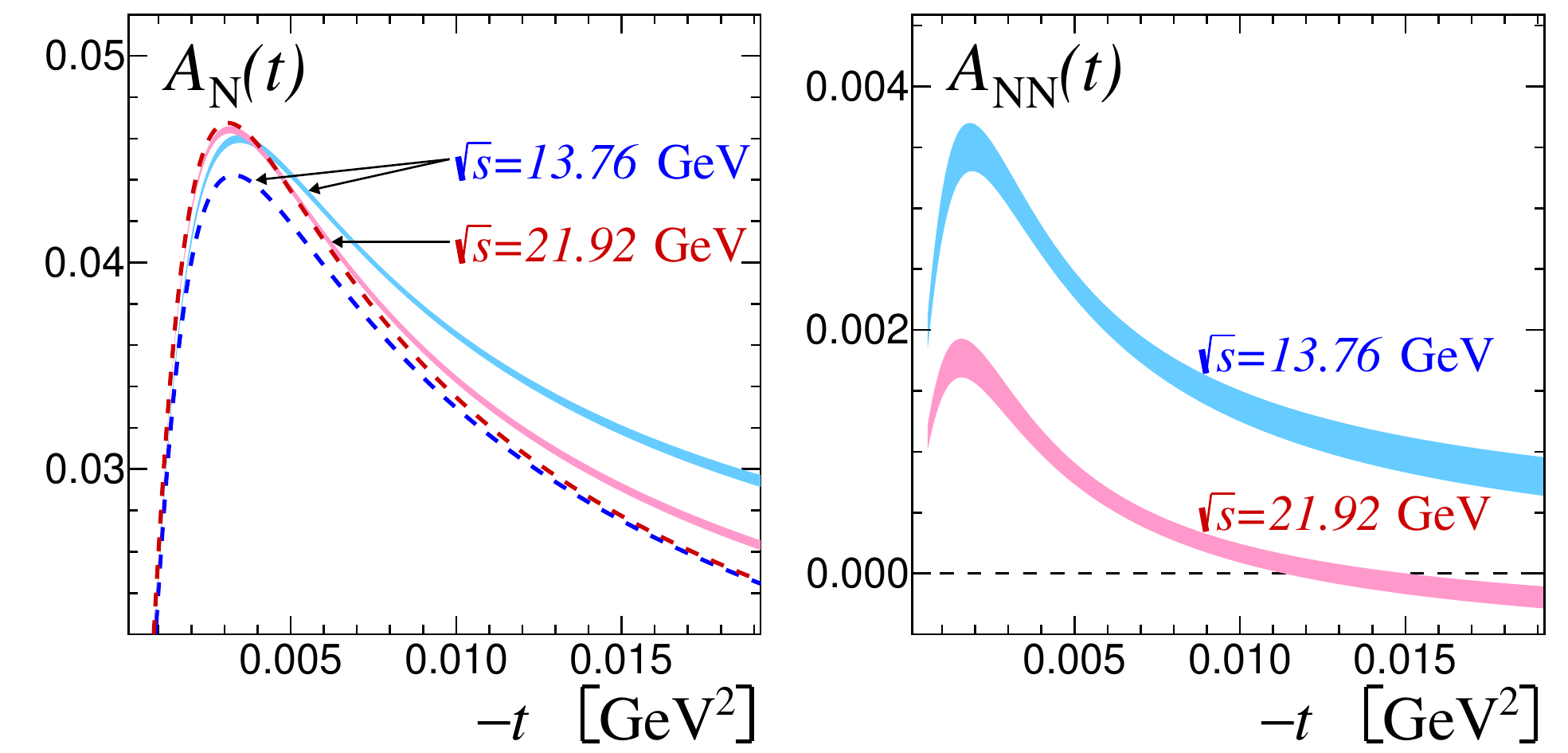}
  \caption{
    Elastic $pp$ analyzing powers $A_\textrm{N}(t)$ and $A_\textrm{NN}(t)$ measured in this work.  The filled areas correspond to $\pm\sigma_\textrm{stat+syst}$. For  $A_\textrm{N}(t)$, the dashed lines refer to the expected analyzing powers if $r_5$\,=\,$0$.
  } \label{fig:AN_ANN}
\end{figure}

For visual control of consistency between the measured single spin asymmetries $a_\textrm{N}^\textrm{j,b}$ and theoretical expectations, it is convenient to use the normalized asymmetry
\begin{equation}
  a_n(T_R) = a_\textrm{N}(t)/A_\textrm{N}(t,r_5\textrm{=0})=P\alpha_5\left(1+\beta_5\,t/t_c\right)
\label{eq:norm}
\end{equation}
which is well approximated by a linear function of $t$ with parameters $\alpha_5(r_5)$\,$\approx$\,$1-2I_5/\varkappa$ and $\beta_5(r_5)$\,$\approx$\,$-2R_5/\varkappa$. The measured $\beta_5$ must be the same for jet and beam asymmetries. The maximum of $A_\textrm{N}(t,r_5$\,=\,$0)$ is about $0.045$ at $T_R$\,=\,$-t/2m_p$\,$\sim$\,$1.7$\,MeV (see Fig.\,\ref{fig:AN_ANN}). 

\begin{figure}[t]
  \includegraphics[width=\columnwidth]{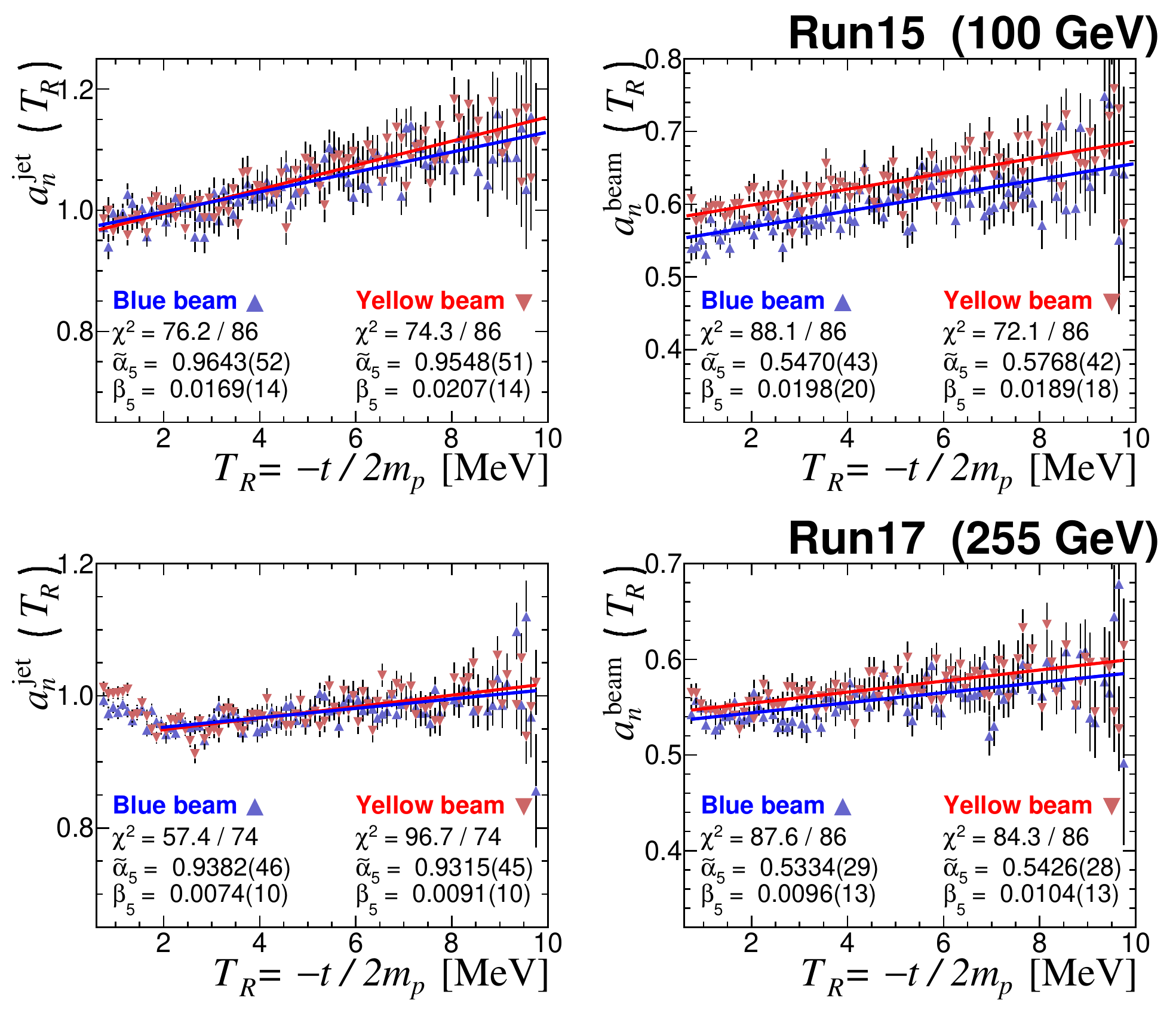} 
  \caption{
    Measured normalized asymmetries in RHIC Run\,15 ($100$\,GeV) and Run\,17 ($255$\,GeV). The fit energy range is $1.9$\,$<$\,$T_R$\,$<$\,$9.9$\,MeV for the $255$\,GeV $a_n^\textrm{jet}(T_R)$ and $0.7\textrm{\,$<$\,}T_R\textrm{\,$<$\,}9.9\,\textrm{MeV}$ for the other graphs. The fit parameter  $\tilde{\alpha}_5$ is defined as $\tilde{\alpha}_5$\,=\,$\langle P\rangle\alpha_5$.
  } \label{fig:Asym}
\end{figure}

Shown in Fig.\,\ref{fig:Asym}, the experimental dependencies $a_n^\textrm{j,b}(T_R)$  are linear functions of $T_R$ in good agreement with expectations. For the $255$\,GeV $a_n^\textrm{jet}(T_R)$, the outlier points at $T_R$\,$<$\,$1.9$\,MeV (presumably due to interference of the magnetic field and inelastic background effects) were eliminated from the data analysis.

An incorrect value of $\rho$ used in the calculation of $A_\textrm{N}(t,r_5\textrm{=}0)$ may result in a false nonlinearity of Eq.\,(\ref{eq:norm}). In the fits with $\rho$ being a free parameter we obtained $\rho$\,=\,$-0.050\pm0.025$ ($100$\,GeV) and $\rho$\,=\,$-0.028\pm0.018$ ($255$\,GeV), values which agree with unpolarized $pp$ data to about 1 standard deviation. So, this test does not indicate any statistically significant discrepancy with the theoretical expectation (\ref{eq:norm}).

To determine the hadronic spin-flip amplitude ratio $r_5$, we fit all four measured asymmetries $a_\textrm{N}^\textrm{j,b}(t)\textrm{=}P_\textrm{j,b}A_\textrm{N}(t,r_5)$ with unknown {\em{}blue} and {\em{}yellow} beam polarizations as free parameters. Nonzero values of $r_5$\,=\,$R_5+iI_5$ were found
\begin{eqnarray}
 \textrm{100 GeV:~}R_5 &=& 
 \left( -16.4 \pm 0.8_\textrm{stat} \pm 1.5_\textrm{syst}\right)\times10^{-3},~~~~~~~{} \label{eq:R5_100} \\
                I_5\,&=&   
 \left( \,\,\,\,-5.3 \pm 2.9_\textrm{stat} \pm 4.7_\textrm{syst}\right)\times10^{-3},  \\ 
\textrm{255 GeV:~}R_5 &=& 
 \left( \,\,\,\,-7.9 \pm 0.5_\textrm{stat} \pm 0.8_\textrm{syst}\right)\times10^{-3}, \\
                I_5\, &=& 
 \left(  {\phantom-}19.4 \pm 2.5_\textrm{stat} \pm 2.5_\textrm{syst}\right)\times10^{-3}. \label{eq:I5_255} 
\end{eqnarray}
The correlation parameters between $R_5$ and $I_5$ are 
$\rho_5^\textrm{stat}$\,=\,$-0.884$, $\rho_5^\textrm{syst}$\,=\,$-0.868$ ($100$\,GeV) and  
$\rho_5^\textrm{stat}$\,=\,$-0.882$, $\rho_5^\textrm{syst}$\,=\,$0.075$ ($255$\,GeV).
The specified systematic errors do not include the effects of  
uncertainties in the external parameters  ($\rho$, $\sigma_\textrm{tot}$, $B$, and $r_p$).
 For both beam energies, the corresponding corrections to $r_5$ can be approximated with sufficient accuracy by  
\begin{eqnarray}
  \Delta R_5  & = & -0.11\times\Delta\rho  
  - \left(0.0019\,\textrm{mb}^{-1} \right)\times\Delta \sigma_\textrm{tot}\nonumber \\
  &+& \left(0.0010\,\textrm{GeV}^{2} \right)\times\Delta B 
  - \left(0.024\,\textrm{fm}^{-1}\right)\times\Delta r_p,~~~~~{} \\
\Delta I_5  & = &  \phantom{-}0.86\times\Delta\rho  
  - \left(0.0085\,\textrm{mb}^{-1} \right)\times\Delta \sigma_\textrm{tot} \nonumber  \\
  &-& \left(0.0011\,\textrm{GeV}^2\right)\times\Delta B. 
\end{eqnarray}
Assessing the values of the external parameters is beyond the scope of this work.

\begin{figure}[t]
  \includegraphics[width=\columnwidth]{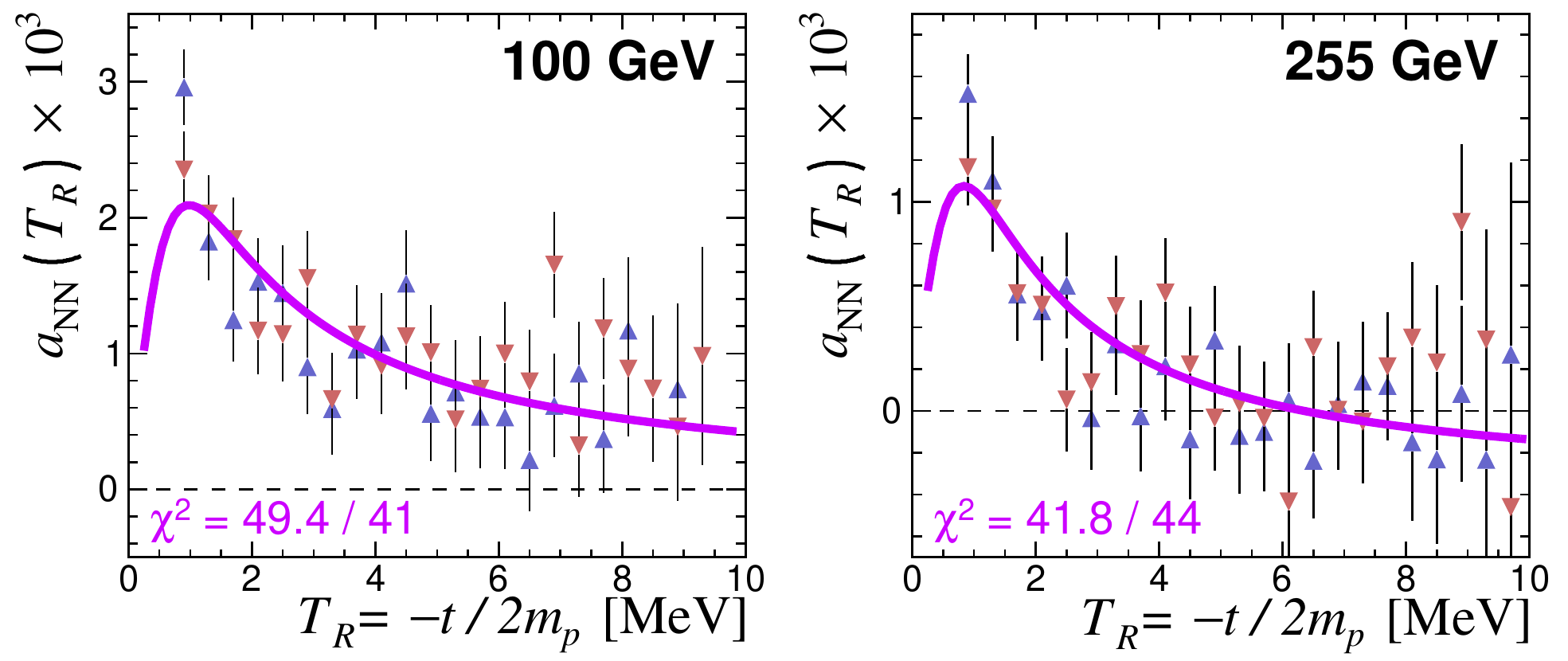}
  \caption{
    Double spin asymmetry $a_\textrm{NN}$ measured at HJET. The fit used values of $P_\textrm{j}$ and $P_\textrm{b}$ from the single spin analysis.
  } \label{fig:ANN_m}
\end{figure}

For the double spin asymmetry $a_\textrm{NN}$ (Fig.\,\ref{fig:ANN_m}), the jet spin correlated systematic uncertainties cancel
 in the ratio $a_\textrm{NN}(T_R)/a_\textrm{N}^\textrm{j}(T_R)$. This statement was verified by comparing the ratio for data with and without background subtraction. Therefore, for the experimental determination of the double spin analyzing power $A_\textrm{NN}(t)$ it is convenient to use the following relation:
\begin{equation}
  A_\textrm{NN}(t) = \frac{A_\textrm{N}(t,r_5)}{\langle P_b\rangle}
  \times \frac{a_\textrm{NN}(t)}{a_\textrm{N}^\textrm{j}(t)}.
  \label{eq:ANN_meas}
\end{equation}
For $r_5$ and $\langle P_b\rangle$ taken from the single spin fit, the experimental uncertainty in (\ref{eq:ANN_meas}) is strongly dominated by the statistical uncertainties of $a_\textrm{NN}(T_R)$:
\begin{eqnarray}
 \textrm{100 GeV:~} & R_2 = & 
 \left( -3.65 \pm 0.28_\textrm{stat} \right)\times10^{-3}, \\
                &I_2 =&  
 \left( -0.10 \pm 0.12_\textrm{stat} \right)\times10^{-3},\\
 \textrm{255 GeV:~} & R_2 = & 
 \left( -2.15 \pm 0.20_\textrm{stat} \right)\times10^{-3},\\ 
                 &I_2 =&  
 \left( -0.35 \pm 0.07_\textrm{stat} \right)\times10^{-3}.
\end{eqnarray}
The correlation parameters are  $\rho_2^\text{stat}$\,=\,$0.860$ ($100$\,GeV) and  $\rho_2^\text{stat}$\,=\,$0.808$ ($255$\,GeV). 
Obviously, non-zero values of $|r_2|$ are well established for both beam energies.

{\em Energy dependence of  $r_5(s)$ and $r_2(s)$.---}%
For unpolarized protons, elastic $pp$ ($\bar{p}p$) scattering can be described 
at low $-t$ with a Pomeron $P$ and the subleading $C$\,=\,$\pm1$  Regge poles for $I$\,=\,$0,1$, encoded by $R^+$ for ($f_2,a_2$) and $R^-$ for ($\omega,\rho$)\,\cite{bib:COMPETE}. In this approach, the unpolarized $pp$ amplitude may be presented as a sum of Reggeon contributions 
\begin{equation}
  \sigma_\textrm{tot}(s)\times\left[\rho(s)+i\right] = \sum_{\mathcal{R}=P,R^\pm}\mathcal{R}(s).
  \label{eq:Regge}
\end{equation}
A basic simple pole approximation assumes 
\begin{equation}
  \mathcal{R}(s)\propto
\left( 1+\zeta_\mathcal{R} e^{-i\pi\alpha_\mathcal{R}} \right)\,
\left(s/4m_p^2\right)^{\alpha_\mathcal{R}-1}
\label{eq:simple}
\end{equation}
with signature factors $\zeta_{R^\pm}$\,=\,$\pm1$, $\zeta_P$\,=\,$+1$ and \lq\lq{standard}\rq\rq intercepts  $\alpha_{R^\pm}$\,=\,$0.5$ and $\alpha_P$\,=\,$1.1$.

\begin{figure}[t]
  \includegraphics[width=\columnwidth]{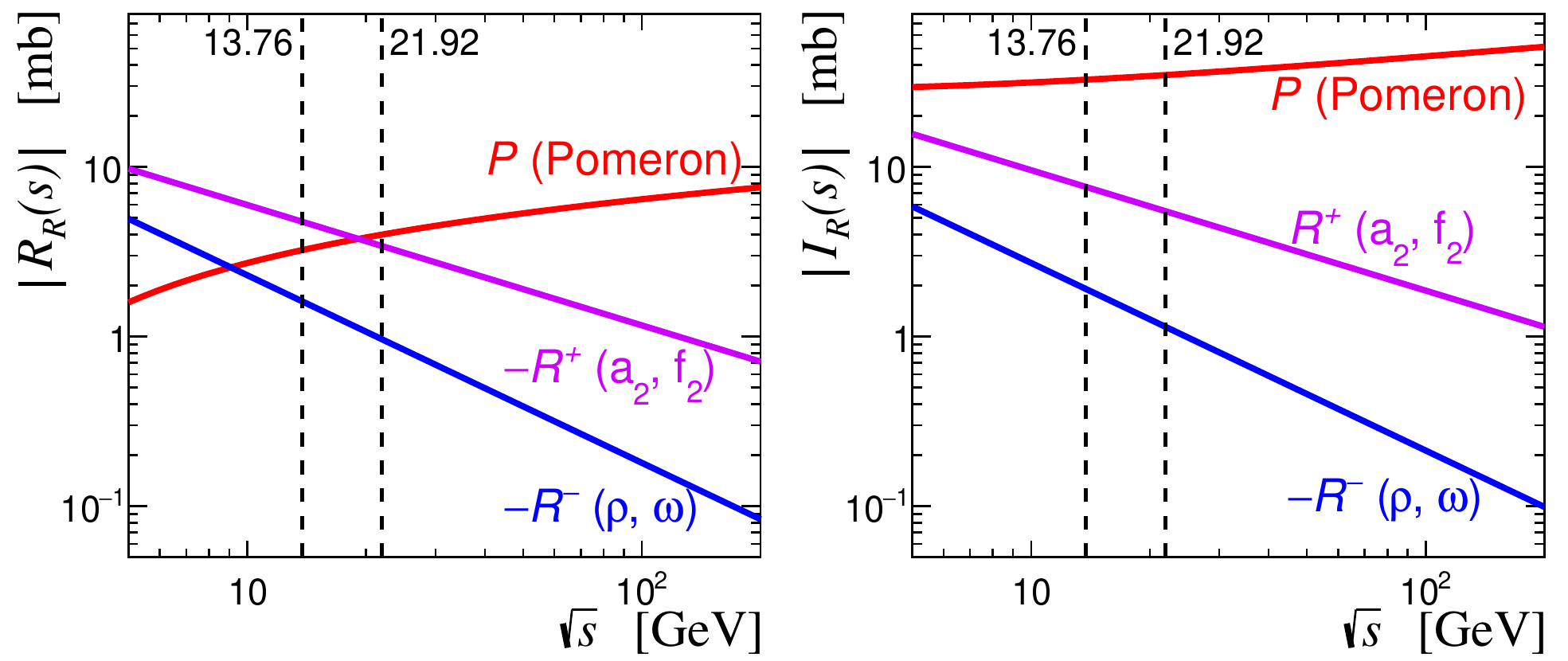}
  \caption{
    The Reggeon contributions $\mathcal{R}(s)$\,=\,$R_\mathcal{R}(s)+iI_\mathcal{R}(s)$ to Eq.\,(\ref{eq:Regge}) defined by the AU-L$\gamma$=2(T) fit of Ref.\,\cite{bib:Menon}.
  } \label{fig:ReIm}
\end{figure}

Here though, we use functions $\mathcal{R}(s)$ as shown in Fig.\,\ref{fig:ReIm} where\,\cite{bib:Menon} the Pomeron is represented by a Froissaron parametrization
\begin{equation}
  P(s)\propto \pi f_F\ln{s/4m_p^2} +i\,\left(1+f_F\ln^2{s/4m_p^2}\right)
  \label{Froissaron}
\end{equation}
with $f_F$\,=\,$0.0090$ and the $R^\pm$ intercepts are $\alpha_{R^+}$\,=\,$0.65$ and $\alpha_{R^-}$\,=\,$0.45$.

In the HJET measurements, $|\textrm{Im}\,r_{5,2}|$ (i.e., both  $|\textrm{Im}\,r_5|$ and  $|\textrm{Im}\,r_2|$) grew with energy indicating that there is a noticeable Pomeron contribution to both single and double spin-flip amplitudes. Moreover, an increasing $|r_5|$ suggests that the Pomeron component dominates in $r_5$ already at HJET energies.

Because of a limited number of the experimental spin-flip entries and following Ref.\,\cite{bib:Trueman}, we expanded $r_{5,2}(s)$ using the above nonflip functions $\mathcal{R}(s)$ scaled by real (because of analyticity in $s$) spin-flip factors $f_{5,2}^\mathcal{R}$
\begin{equation}
  \sigma_\textrm{tot}(s)\times r_{5,2}(s) = \sum_{\mathcal{R}=P,R^\pm}{f_{5,2}^\mathcal{R}\,\mathcal{R}(s)}.
  \label{eq:r5,2}
\end{equation}

In a combined fit of the $100$ and $255$\,GeV HJET data, we find
\begin{eqnarray}
  f_5^{P^{\phantom+}} & = & {\phantom-}0.045 \pm 0.002_\textrm{stat} \pm 0.003_\textrm{syst},
  \label{eq:f5P}\\
  f_5^{R^+}                 & = & -0.032\pm 0.007_\textrm{stat} \pm 0.014_\textrm{syst}, \label{eq:f5R+}\\
  f_5^{R^-}                 & = & {\phantom-}0.622\pm 0.023_\textrm{stat} \pm 0.024_\textrm{syst}.
  \label{eq:f5R-} 
\end{eqnarray}

Similarly, for the double spin-flip amplitude expansion we obtain
\begin{eqnarray}
  f_2^{P\phantom{+}}   &=&  -0.0020 \pm 0.0002_\textrm{stat}, \label{eq:f2P}\\ 
  f_2^{R^+} &=&  \phantom{-}0.0162\pm 0.0007_\textrm{stat}, \label{eq:f2R+}\\
  f_2^{R^-} &=&  \phantom{-}0.0297\pm 0.0041_\textrm{stat}. \label{eq:f2R-}
\end{eqnarray}

  At high energies where the contributions $R^\pm$ decay, the model (\ref{eq:r5,2}) used gives the following spin-flip parameters:
\begin{equation}
  r_{5,2}(s)=f_{5,2}^P\times\left[\rho(s)+i\right].
\end{equation}

In terms of the Pomeron anomalous magnetic moment introduced in Ref.\,\cite{bib:KZ89}, the fit yields  $M_{I\!\!P}$\,=\,$2f_5^P$\,=\,$0.09\pm0.01$. The provisional value of $r_P$$\,\sim\,$$0.03$\,\cite{bib:BKLST} derived from $\pi p$  data\,\cite{bib:Borghini} at $6\textendash14$\,GeV/$c$ can, using assumption (\ref{eq:r5,2}), be related to $f_5^P$\,$\approx$\,$r_P$ in reasonable agreement with Eq.\,(\ref{eq:f5P}).

The value of $f_5^P$\,=\,$0.10\pm0.01$\,\cite{bib:Trueman} estimated from $p^\uparrow\textrm{C}$ data is noticeably larger than in Ref.\,(\ref{eq:f5P}). However, this estimate required a model dependent conversion from proton-nucleus asymmetries to proton-proton $r_5$ and, also, was strongly based on unpublished experimental results\,\cite{bib:pC} with undetermined systematic uncertainties.

\begin{figure}[t]
  \includegraphics[width=\columnwidth]{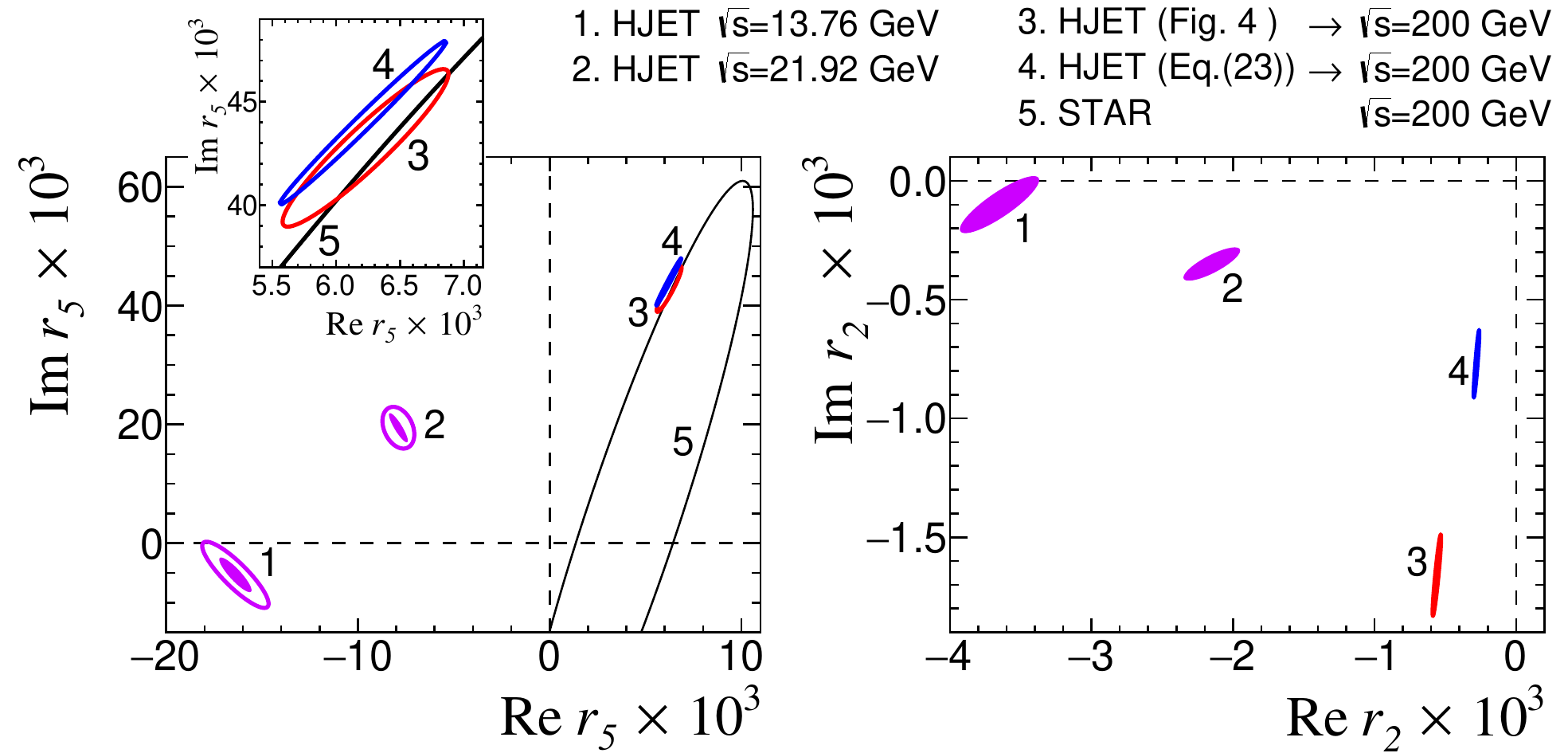}
  \caption{
    $\Delta\chi^2$\,=\,$1$ correlation (stat+syst) contours for $r_5$ and $r_2$. Filled ellipses mean statistical error only. The HJET extrapolations to 200\,GeV are labeled 3 and 4. The STAR Collaboration result\,\cite{bib:STAR} for $r_5$ was changed by us using Eqs.\,(\ref{eq:tc'})--(\ref{eq:kappa'}).
  } \label{fig:r5_r2}
\end{figure}

The $r_5(s)$ and $r_2(s)$ dependencies on the beam energy are illustrated in Fig.\,\ref{fig:r5_r2} where the extrapolations to $\sqrt{s}$\,=\,$200$\,GeV, based on the Froissaron parametrization (\ref{Froissaron}), are labeled \lq\lq3.\rq\rq  Consistency between the extrapolation of $r_5$  and the STAR Collaboration measurement\,\cite{bib:STAR} was observed, though the STAR experimental uncertainties are not inconsiderable.
 
It is interesting to note that the values of $r_5$ and $r_2$, when projected from $\sqrt{s}$ $14\textendash22$ to $200$\,GeV, have smaller uncertainties than those of the measurements. This may be explained by decay of the $R^\pm$ pole contributions at large $s$ and by using functions $\mathcal{R}(s)$ that are too tightly constrained (which, for the selected model, is a good approximation in the energy range considered). However, many models\,\cite{bib:COMPETE} are used to parametrize $\sigma_\textrm{tot}(s)$ and $\rho(s)$ which may render $\mathcal{R}(s)$ more uncertain.

To estimate the dependence of a Reggeon analysis on a particular model, we also fitted the HJET data using a sum of simple poles (\ref{eq:simple}). These extrapolations of $r_5$ and $r_2$ to $\sqrt{s}$\,=\,$200$\,GeV are labeled \lq\lq4\rq\rq in Fig.\,\ref{fig:r5_r2}. Since, at HJET energies, the double spin-flip amplitude is dominated by an $R^+$ contribution, the $r_2$ projection to $200$\,GeV is strongly affected by a  variation of $\alpha_{R^+}$.

The expansions (\ref{eq:r5,2}) fit the measurements with statistically insignificant discrepancies $\chi^2$\,=\,$2.2$ [Eqs.\,(\ref{eq:f5P})--(\ref{eq:f5R-})] and   $\chi^2$\,=\,$1.6$ [Eqs.\,(\ref{eq:f2P})--(\ref{eq:f2R-})] for ndf=1 showing consistency between the experimental data and Eq.\,(\ref{eq:r5,2}).

To evaluate a possible difference between single spin-flip (sf) and nonflip functions $P(s)$, we determined the ratio $\tilde{f}_F^\textrm{(sf)}$\,=\,$f_F^\textrm{(sf)}/f_F$  in a combined analysis including the STAR Collaboration result. For a fixed  $\alpha_{R^+}$\,=\,$0.65$, we obtained $\tilde{f}_F^\textrm{(sf)}$\,=\,$0.5\pm0.5$ and $\alpha_{R^-}^\textrm{(sf)}$\,=\,$0.62\pm0.11$. However, $\tilde{f}_F^\textrm{(sf)}$ strongly depends on the $\alpha_{R^+}$ selection. The fit of the  Pomeron spin-flip intercept [using a simple pole for $P(s)$] is stable in a wide range of  $0.3$\,$<$\,$\alpha_{R^+}$\,$<$\,$0.8$. It gives
\begin{equation}
\Delta_P^\textrm{(sf)}=\alpha_P^\textrm{(sf)}-1=0.117\pm0.031_\textrm{~stat+syst},
\end{equation}
 which agrees with the unpolarized $\Delta_P$\,=\,$0.096^{+0.012}_{-0.009}$\,\cite{bib:Cudell}, and $\alpha_{R^-}^\textrm{(sf)}$\,=\,$0.65\!\pm\!0.11$.

{\em Summary.---}%
In RHIC polarized proton runs 2015 (100\,GeV) and 2017 (255\,GeV), we have measured elastic $pp$ analyzing powers in the CNI region $0.0013$\,$<$\,$-t$\,$<$\,$0.018$\,GeV  with accuracy $|\delta A_\textrm{N,NN}(t)|$\,$\sim$\,$2\times10^{-4}$\,\cite{bib:SPIN18} as shown in Fig.\,\ref{fig:AN_ANN}. To graph $A_N(t)$, we substituted the fitted values of $r_5$ from Eqs.\,(\ref{eq:R5_100})--(\ref{eq:I5_255}) in Eq.\,(\ref{eq:AN}), taking into account statistical and systematic uncertainties and their covariances. In fact, this is equivalent to determining $A_N(t)$ directly from the linear fit of the normalized asymmetries $a_n(T_R)$. Thus, the result is not greatly affected by absorptive corrections, nor by possible variations in $\rho$, $\sigma_\textrm{tot}$, $B$, and $r_p$.

The accuracy achieved in the determination of $A_N(t)$ allows one to use a higher density {\em{}unpolarized} hydrogen jet target in a high precision absolute polarimeter, e.g.,~at a future EIC\,\cite{bib:EIC}. For a 30-fold increase in jet density, the expected statistical and systematic uncertainties of the polarization measurement would be $\delta^\textrm{stat}P$\,$\lesssim$\,$1\%/$h and $\delta^\textrm{syst}P/P$\,$\lesssim$\,$1\%$.

The hadronic spin-flip amplitude ratios $r_5$ and $r_2$ were reliably  isolated at both energies. 
Applying the corrections indicated in Eqs.\,(\ref{eq:tc'})--(\ref{eq:kappa'}) to the expression\,\cite{bib:BKLST} for  $A_\textrm{N}(t)$  resulted in a change of the measured $r_5$ by about the size of the experimental uncertainty. The absorptive corrections were not included in the data analysis, but, if they become available, a simple correction to Re\,$r_5$ could be applied.

Measurements at two energies permitted a Regge pole analysis of elastic $pp$ scattering to be extended to the spin dependent case. A Reggeon expansion of the spin-flip parameters $r_5(s)$ and $r_2(s)$ indicated that Pomeron single and double spin-flip couplings were well determined and found to be significantly different from zero. However, the absorptive corrections when available, might require a re-analysis of the expansion.

\begin{acknowledgments}
We thank the Collider Accelerator Department and the RHIC/AGS Operation Groups. We also would like to thank A.~Bazilevsky, B.Z.~Kopeliovich, and M.~Krelina for useful discussions. B.Z.~Kopeliovich read the manuscript and made valuable comments. This work was supported by Brookhaven Science Associates, LLC under Contract No.~DE-AC02-98CH\,10886 with the U.S. Department of Energy. Funding was also provided from the RIKEN BNL Research Center. N.H.B. is grateful for partial support from the School of Mathematics.
\end{acknowledgments}

\end{document}